\documentclass[a4paper]{jpconf}
\usepackage{graphicx}
\begin{document}
\title{$B_s$ Mixing and $B$ Hadron Lifetimes at CDF}

\author{Michael Milnik on behalf of the CDF Collaboration}

\address{Institut f\"ur Experimentelle Kernphysik, Universit\"at Karlsruhe, Wolfgang-Gaede-Str. 1, 76131 Karlsruhe, Germany}

\ead{milnik@ekp.physik.uni-karlsruhe.de}

\newcommand{\Bs}{B_{s}^0} 
\newcommand{\Bsb}{\bar{B}_{s}^0} 
\newcommand{\Bc}{B_{c}} 
\newcommand{\Bu}{B_{u}^+} 
\newcommand{\Bd}{B_{d}^0} 

\begin{abstract}
We present the CDF results using 1.0 fb$^{-1}$ of data on the mixing frequency measurement in the $\Bs$ system and the lifetime measurements of several $B$ hadrons. We obtain
$\Delta m_s=17.77\pm 0.1\pm 0.07 $ ps$^{-1}$ and $c\tau(\Lambda_b)=473.8 \pm 23.1 \pm 3.5$ $\mu$m. The later one is more than 3$\sigma$ above the world average, but in reasonable agreement with HQE calculations.\end{abstract}

\section{Introduction}
The Collider Detector at Fermilab (CDF) is placed at one of the interaction points of the Tevatron, a $p\bar{p}$ collider at a center of mass energy of $\sqrt{s}=1.96$ TeV. This energy allows to produce and study the properties of $B$ mesons and baryons.

Measuring these properties allows us to determine fundamental parameters and constrain the parameter space of models of physics beyond the Standard Model. We focus here on $\Bs$ mixing and the $\Lambda_b$ lifetime measurements, since these are unique to the Tevatron.

\section{$\Bs$ Mixing}
A $\Bs$ meson produced at time $t=0$ can oscillate into its anti-particle $\bar{B}_s$ via second order weak processes. The largest contribution to the $\Bs-\Bsb$ oscillation comes from the virtual top quark contribution. The measurement of the mixing frequency in the $\Bs$ system allows to constrain the CKM matrix element $V_{ts}$ and thus constrain one of the fundamental parameters of the Standard Model. 

Measuring the oscillation frequency of the $\Bs$ systems is a complex task, since several properties of the decay have to be determined. 

\begin{itemize}

\item The decay flavor is determined using flavor specific modes. Several channels are combined to maximize the signal yield: the $\Bs$ is reconstructed by $D_s\pi$, $D_s3\pi$ and $D_sl^+X$ decays, where the $D_s$ can decay to $\phi\pi$, $K^*K$ or $3\pi$. Also partially reconstructed decay channels are used. The events were triggered by the Two-Track-Trigger and the selection is optimized using neural networks.

\item  A good decay length resolution is very important to resolve the fast oscillation frequency in the $\Bs$ system. The innermost layer of the silicon detector at CDF is mounted directly on the beam pipe, allowing for a position measurement closer than 2 cm to the primary vertex.

\item Several flavor tagging techniques are used to estimate the flavor at production time. On the same side as the reconstructed meson, the kaon from fragmentation in the vicinity of the $\Bs$ meson can be used.

On the opposite side the information of several tagging algorithms is combined using a neural network: the charge of the jet associated with the other $b$ quark from the $b\bar{b}$ production, identified leptons from $b\to l^-X$ decays and kaons from the cascade $b\to c\to K^-X$.

\end{itemize}

\begin{figure}
\hspace{0.3cm}
\includegraphics[width=0.55\textwidth]{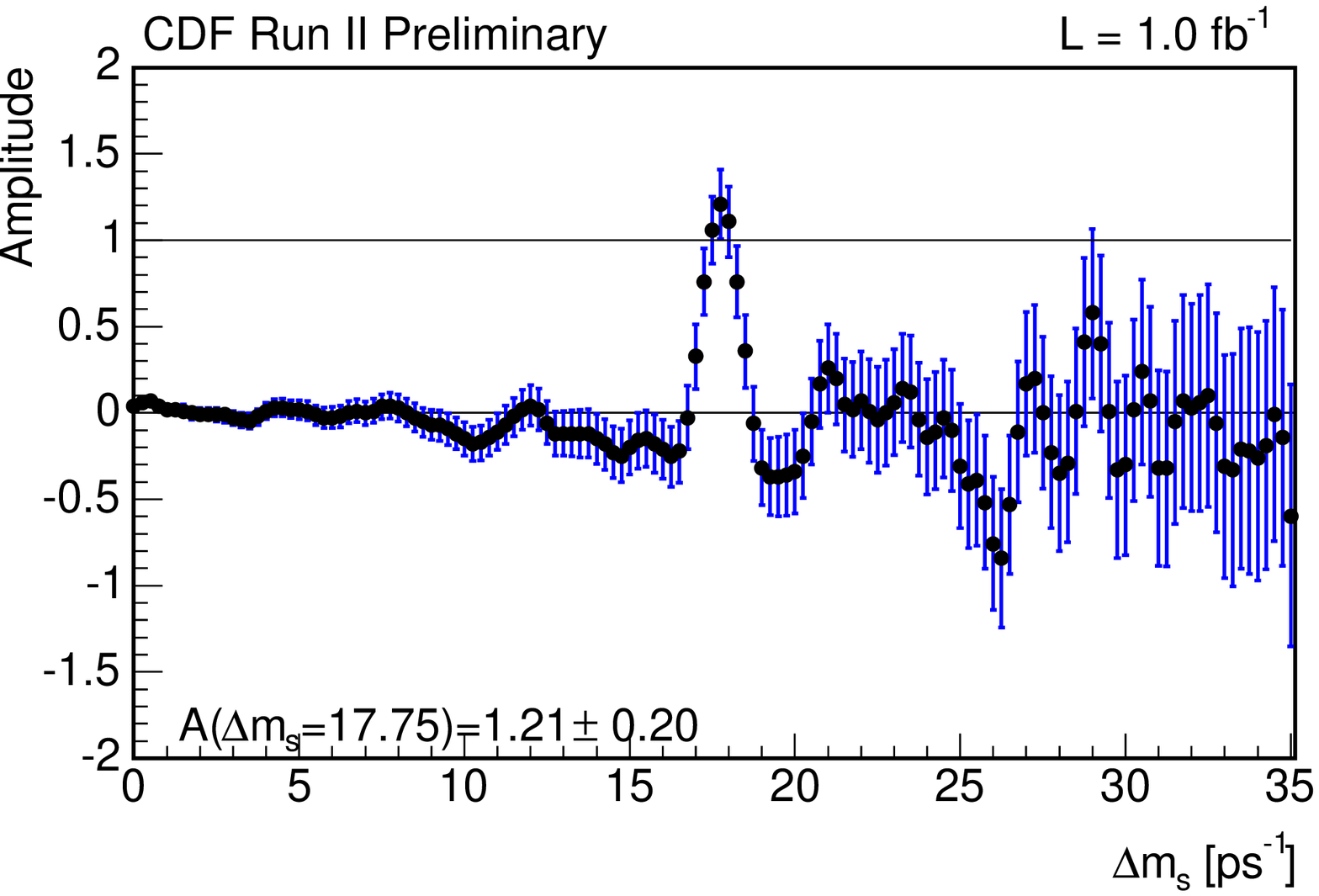}
\includegraphics{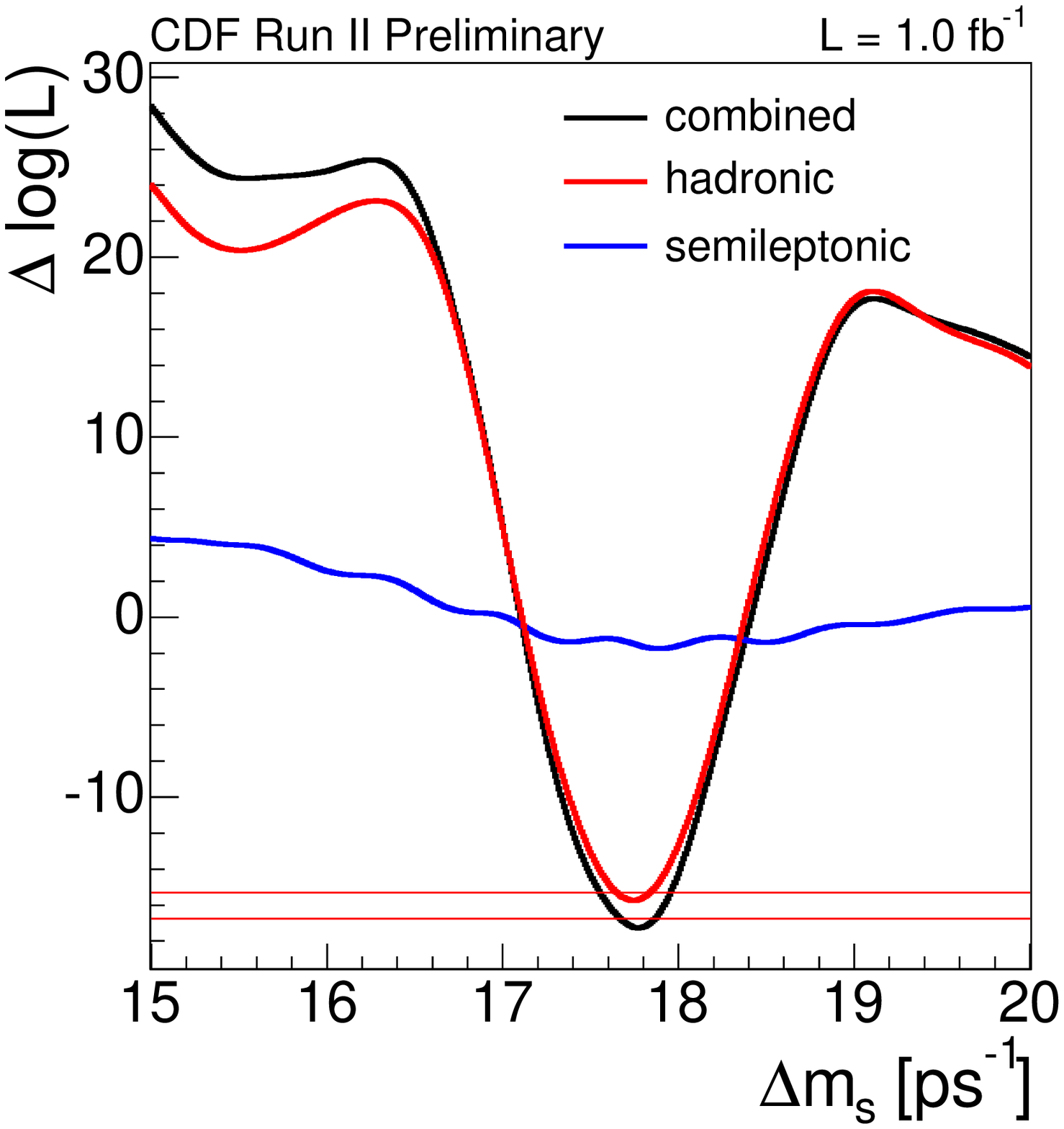}
\caption{\label{fig:ampscan}Amplitude scan of the mixing frequency and likelihood scan of the final result.}
\end{figure}

The likelihood describing the proper decay time of the tagged $\Bs$ meson can be written as $P(t,\sigma_t) \propto 1+ \xi AD \cos(\Delta m_s t)$,
where the amplitude $A$ is fitted with a fixed mixing frequency $\Delta m_s$, $\xi$ is the tagging decision and $D$ its estimated dilution. Scanning through different frequencies, $A$ should be one at the true frequency, zero otherwise. Figure \ref{fig:ampscan}(left) shows the result of the amplitude scan. The final result is obtained by an unbinned maximum likelihood fit that measures $\Delta m_s=17.77\pm 0.1\pm 0.07$ ps$^{-1}$ with a significance $>5\sigma$\cite{Abulencia:2006ze}. Figure \ref{fig:ampscan}(right) shows the likelihood destribution around the minimum.

This measurement allows to calculate the CKM matrix element ratio $V_{td}/V_{ts}=0.2060\pm 0.0007$(exp.)$ ^{+0.0081}_{-0.0060}$(theo.), in which many theoretical uncertainties cancel out. 
Since the CKM matrix elements $V_{ts}$ and $V_{cb}$ are the same in the Wolfenstein approximation, the ratio can be used to constrain the angle $\beta$ of the unitary triangle.

\vspace{-0.08cm}
\section{$B$ Hadron Lifetimes}

In principle, the measurement of the weak $b \to c(u)$ quark decay via the emission of an $W$ boson allows to determine the associated CKM matrix elements. Since there are no free quarks, we have to analyze bound states of a $b$ quark with lighter quarks. In the simple spectator model, the lifetime of the $B$ hadrons is dominated by the $b$ quark decay and implies that all $B$ hadrons have the same lifetime. However, this does not agree with the observed lifetime hierarchy: $\tau_{\Bc} < \tau_{\Lambda_b} <\tau_{\Bs} \approx \tau_{\Bd} < \tau_{\Bu}$.

The interaction between the quarks in the bound state changes the lifetime of the hadron. This is calculated using the Heavy Quark Expansion (HQE)\cite{Bigi:1995jr,Beneke:2002rj}.
Qualitatively HQE does explain the observed hierarchy of the $B$ hadron lifetimes, and most quantitative findings agree very well. Measuring $\tau_{\Lambda_b}$ allows to check the current theoretical predictions.

Lifetimes of several $B$ hadrons with similar decay topology are measured at CDF. Measurements by the $B$ factories allow to determine very precisely the lifetimes of the lightest $B$ hadrons: $\Bu$ and $\Bd$. The heavier hadrons are only produced at the Tevatron\footnote{Except at a test run at Belle on the $Y(5S)$ resonance.}. The $\Bd$ and $\Bu$ lifetime measurements can therefore be used to cross check the analysis strategy. 

The following decays are reconstructed and their lifetime is measured: \newline
\begin{center}
\begin{tabular}{l|l}
Decay Channel & $\tau$ $\pm$ stat. $\pm$ syst. $[$ps$]$ \\ \hline 
$\Bd \to J/\psi K_s$, $\Bd \to J/\psi K^*$& $1.551 \pm 0.019 \pm 0.011$ \\
$\Bu \to J/\psi K^+$ &$1.630 \pm 0.016 \pm 0.011$ \\
$\Bs \to J/\psi \phi$ &$1.494 \pm 0.054 \pm 0.009$ \\
$\Lambda_b \to J/\psi \Lambda$ &$1.580 \pm 0.077 \pm 0.012$, \\
\end{tabular}
\end{center}
\vspace{0.1cm}
where the $J/\psi \to \mu \mu$ decay is used to trigger the events. For the reconstruction of the $B$ hadron decay vertex, only the di-muon vertex is used since the $\Lambda$ and $K_s$ are longlived particles which could introduce a bias to longer lifetimes.

\begin{figure}
\begin{center}
\includegraphics[width=0.45\textwidth]{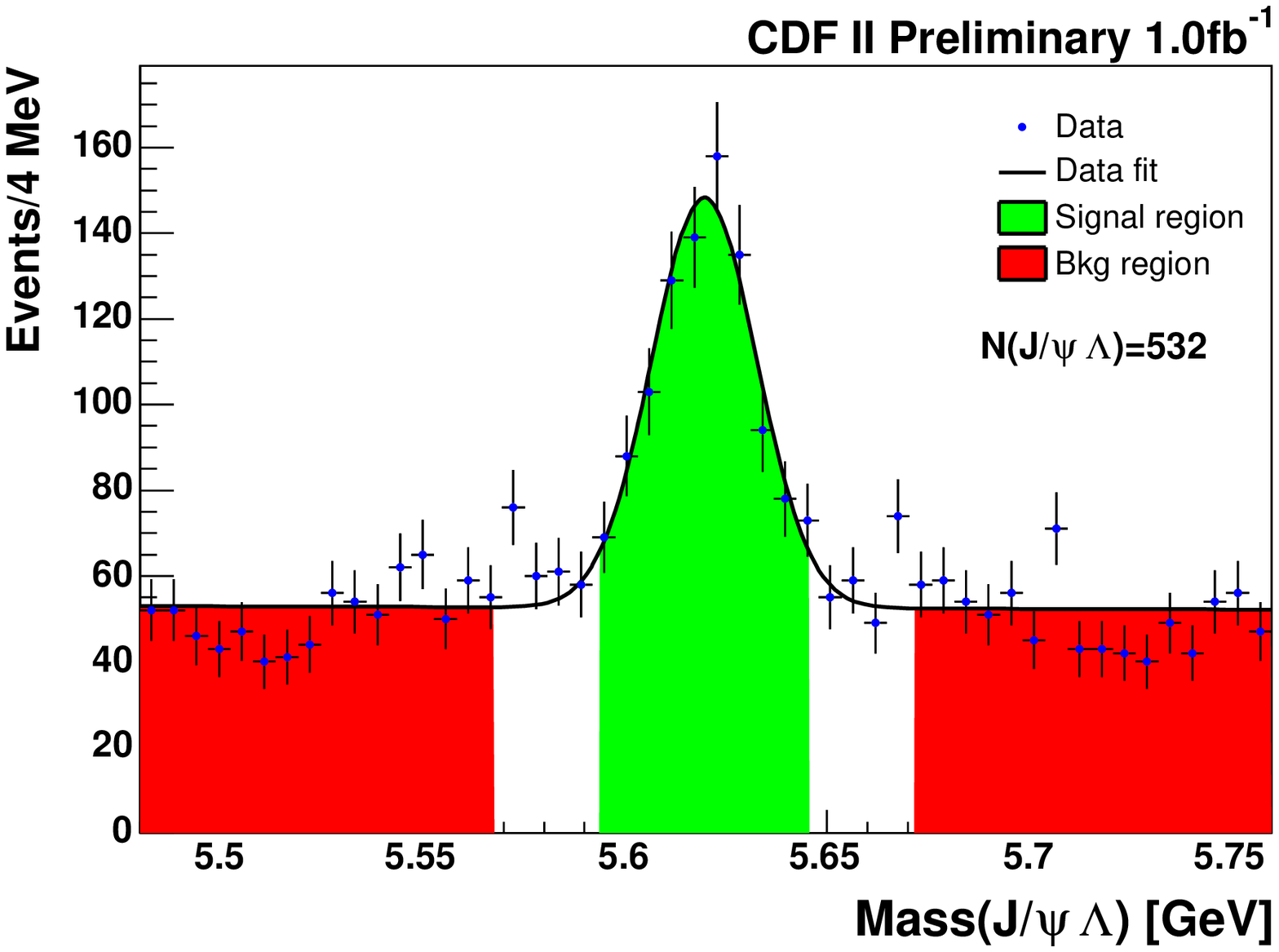}
\includegraphics[width=0.45\textwidth]{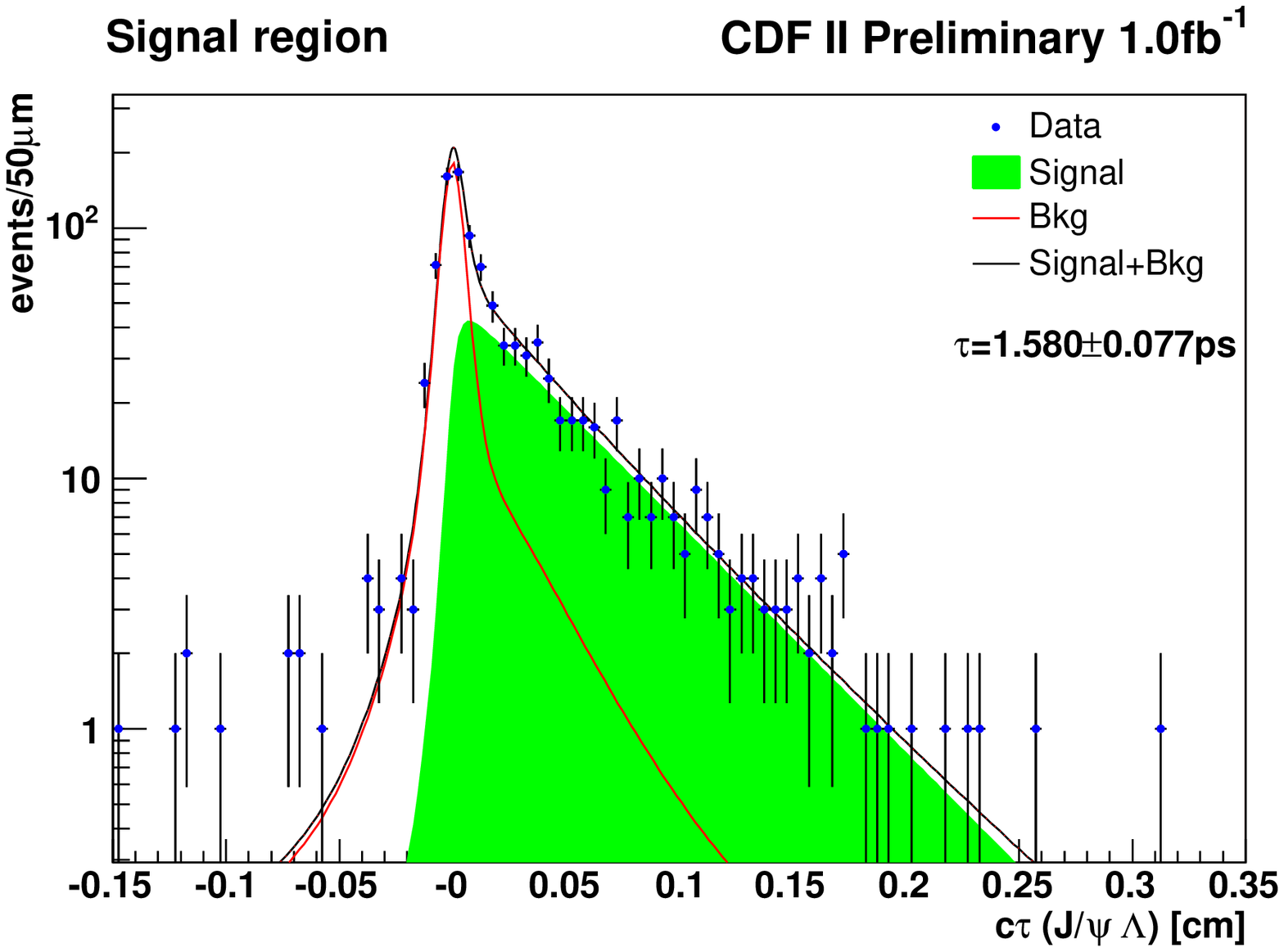}
\end{center}
\caption{\label{fig:lampro}$\Lambda_b$ mass and lifetime projections.}
\end{figure}

The measurements of the $\Bu$, $\Bd$ and $\Bs$ lifetimes are in good agreement with the world average and theoretical predictions. The fit projections of the mass and lifetime for the $\Lambda_b$ decay are shown in Figure \ref{fig:lampro}. The obtained signal yield, using 1.0 fb$^{-1}$, is $N(J/\psi \Lambda)=532$ and we measure a lifetime of $c\tau(\Lambda_b)=473.8 \pm 23.1 \pm 3.5$ $\mu$m\cite{Abulencia:2006dr}.

This result deviates 3$\sigma$ from the current world average. Whereas all previous measurements are below the theoretical prediction of HQE, the CDF result is above. Using the obtained numbers we can calculate the lifetime ratio of $\tau (\Lambda_b)/\tau (\Bu)=1.018 \pm 0.062\pm 0.007$, which compares to the theoretical prediction of $\tau (\Lambda_b)/\tau (\Bu)=0.88 \pm 0.05$. 

\section{Conclusion and Outlook} 
The Tevatron is an excellent place for $B$ hadron studies. Using 1.0 fb$^{-1}$ of data, CDF is able to put theoretical predictions to stringent tests. CDF provides the first observation of $\Delta m_s$ and very precise $B$ hadron lifetime measurements. Having already more then 2.5 fb$^{-1}$ on tape, with prospect for more, the heavy flavor physics program at the Tevatron will produce more precise results in the future.

\section*{References}

\bibliographystyle{unsrt}\flushleft
\bibliography{bib}

\end{document}